\documentstyle[12pt]{article}

\newcommand{\be}{\begin{equation}}
\newcommand{\ee}{\end{equation}}
\newcommand{\ba}{\begin{eqnarray}}
\newcommand{\ea}{\end{eqnarray}}

\begin{document}
\begin{flushright}
JINR preprint E2-98-348
\end{flushright}
\begin{center}
{\bf
\Large
{Nambu--Poisson reformulation
 of the finite dimensional dynamical systems }}
\end{center}
\begin{center} Dumitru Baleanu\footnote{ Permanent address
: Institute of Space Sciences, P.O.BOX, MG-36, R
76900,Magurele-Bucharest, Romania,E-Mail
address:~~baleanu@thsun1.jinr.ru,baleanu@venus.ifa.ro}\\ Bogoliubov
Laboratory of Theoretical Physics \\ Joint Institute for Nuclear
Research\\ Dubna, Moscow Region, Russia
\end{center}
\begin{center}
    and
\end{center}
\begin{center}
 Nugzar Makhaldiani\footnote{e-mail address:~~mnv@cv.jinr.ru}
\end{center}
\begin{center}
Laboratory of Computing Techniques and Automation\\
Joint Institute for Nuclear Research\\
 Dubna, Moscow Region, Russia
\end{center}

\vskip 5mm
\bigskip
\nopagebreak
\begin{abstract}
In this paper we introduce a system of nonlinear ordinary
differential equations
which in a particular case reduces to Volterra's system.
We found in two simplest cases the complete sets of the integrals of motion
using Nambu--Poisson reformulation of the Hamiltonian dynamics.
In these cases we have solved the systems by quadratures.
\end{abstract}

{\bf 1.} Introduction\\

       The Hamiltonian mechanics (HM) is in the ground of
mathematical description of the physical theories, \cite{fad}. But HM is in
a sense blind, e.g., it does not make difference between two opposites: the
ergodic Hamiltonian systems (with just one integral of motion) and
integrable Hamiltonian systems (with maximal number of the integrals
of motion).\\

  By our proposal, Nambu's mechanics (NM) \cite{nambu} is proper generalization
of the HM, which makes difference between dynamical systems with different
numbers of integrals of motion explicit.\\

 In this paper we introduce a system of nonlinear ordinary
differential equations
which in a particular case reduces to Volterra's system, \cite{volterra}
and integrate this system using Nambu--Poisson formalism, \cite{nambu,leon}.

In Sec.2 of  this paper we introduce the dynamical system.
In Sec.3 and Sec.4  we construct a complete set of integrals
of motion in two particular cases for which we found the general solutions
in quadratures. In Sec.5 we found some integrals of motion in the general
case and present our conclusions.\\

{\bf 2.} \ The  system\\

In this section we introduce the following dynamical system

\ba\label{din}
&&{\dot x_{n}}=\gamma_{n}\sum_{m=1}^{p}(e^{x_{n+m}}-e^{x_{n-m}}), \cr
&&{1\leq n\leq N}, \ {1\leq p}\leq[{(N-1)/2}],\ 3\leq N, \cr
&& x_{n+N}=x_{n},
\ea
where $\gamma_{n}$ are real numbers, and $[a]$ means
the integer part of a.\\
The system, (\ref{din}) for $\gamma_{n}=1$, $p=1$ and $x_n=lnv_n$,
becomes Voltera's system
\be
{\dot v_{n}}=v_{n}(v_{n+1}-v_{n-1}),
\ee
then it is connected also to the Toda's lattice
system, \cite{toda}
\ba
{\dot y_{n}}=e^{y_{n+1}-y_{n}} +e^{y_{n}-y_{n-1}}.\nonumber
\ea

Indeed if
\ba x_{n}=y_{n}-y_{n-1},\nonumber
\ea
then
\ba
{\dot x_{n}}=e^{x_{n+1}}-e^{x_{n-1}}.\nonumber
\ea

If $\gamma_n=1$ and ${p\ge 1}$,
 the system (\ref{din}) reduces to
the  so-called Bogoiavlensky lattice system, \cite{bogo}
\be\label{bo}
{\dot v_{n}}=v_{n}\sum_{m=1}^{p}(v_{n+m}-v_{n-m}).
\ee

For $N=3$, $p=1$ and arbitrary $\gamma_{n}$, (\ref{din}) is connected
to the system of three vortexes of two-dimensional
ideal hydrodynamics, \cite{vort,mkh}.\\

{\bf 3.} The case of $N=3, p=1$\\

 It is well known  that the system of $N$ vortexes can be
described by the following system of differential equations, \cite{vort}
\begin{equation}
\dot z_n=i\sum_{m\not=n}^{N}\frac{\gamma_m}{z_n^*-z_m^*},
\end{equation}
where $z_n=x_n+iy_n$ are complex coordinate of the centre of n-th vortex.

For $N=3,$ it is easy to verify that the quantities
\begin{eqnarray}
&&x_1=ln|z_2-z_3|^2, \\ \nonumber
&&x_2=ln|z_3-z_1|^2, \\ \nonumber
&&x_3=ln|z_1-z_2|^2
\end{eqnarray}
satisfy the following system
\ba\label{vor}
{\dot x_{1}}=\gamma_{1}(e^{x_{2}}-e^{x_{3}}),\cr
{\dot x_{2}}=\gamma_{2}(e^{x_{3}}-e^{x_{1}}),\cr
{\dot x_{3}}=\gamma_{3}(e^{x_{1}}-e^{x_{2}}),
\ea
after change of the time parameter as
\begin{equation}
dt=\frac{e^{(x_1+x_2+x_3)}}{4S}d\tau=e^{(x_1+x_2+x_3)/2}Rd\tau,
\end{equation}
where $S$ is the area of the triangle with vertexes in the
centres of the vortexes and $R$ is the radius of the circle with the vortexes
on it.

 The system (\ref{vor}) has two integrals of motion
\begin{eqnarray}\label{hami}
&&H_{1}=\sum_{i=1}^{3}\frac{e^{x_i}}{\gamma_i}, \\  \nonumber
&&H_{2}=\sum_{i=1}^{3}\frac{x_i}{\gamma_i}
\end{eqnarray}
\noindent
and can be presented in the Nambu--Poisson form, \cite{mkh}
\begin{eqnarray}\label{nambu1}
&&\dot x_{i}=\omega_{ijk}\frac{\partial H_1}{\partial x_j}
\frac{\partial H_2}{\partial x_k} \\  \nonumber
&&=\{x_{i},H_1,H_2\}=\omega_{ijk}\frac{e^{x_j}}{\gamma_{j}}
\frac{1}{\gamma_{k}},
\end{eqnarray}
where
\begin{eqnarray}
&&\omega_{ijk}=\epsilon_{ijk}\rho,\\  \nonumber
&&\rho=\gamma_{1}\gamma_{2}\gamma_{3}  \nonumber
\end{eqnarray}
and the Nambu--Poisson bracket of the functions $A,B,C$ on the
three-dimensional phase space is
\begin{eqnarray}\label{nambu2}
\{ A,B,C\}=\omega_{ijk}\frac{\partial A}{\partial x_i}\frac{\partial B}
{\partial x_j}\frac{\partial C}{\partial x_k}.
\end{eqnarray}

The fundamental bracket is
\begin{eqnarray}\label{nambu3}
\{ x_1,x_2,x_3\}=\omega_{ijk}.
\end{eqnarray}

Then we can again change the time parameter as
\begin{eqnarray}
du=\rho d\tau
\end{eqnarray}
and obtain Nambu's mechanics, \cite{mkh}
\begin{eqnarray}
\dot x_{i}=\epsilon_{ijk}\frac{\partial H_1}{\partial x_j}
\frac{\partial H_2}{\partial x_k}.    \nonumber
\end{eqnarray}

{\bf 4.} The next important case  is  $N=4$ and $p=1$,
\ba\label{system}
{\dot x_{1}}=\gamma_{1}(e^{x_{2}}-e^{x_{4}}),\cr
{\dot x_{2}}=\gamma_{2}(e^{x_{3}}-e^{x_{1}}),\cr
{\dot x_{3}}=\gamma_{3}(e^{x_{4}}-e^{x_{2}}),\cr
{\dot x_{4}}=\gamma_{4}(e^{x_{1}}-e^{x_{3}}).
\ea

 Like as $N=3,p=1$ case, for (\ref{system}) we have two integrals of motion
\be\label{h1}
H_{1}={e^{x_{1}}\over\gamma_{1}}
+{e^{x_{2}}\over\gamma_{2}} +
{e^{x_{3}}\over\gamma_{3}}+
{e^{x_{4}}\over\gamma_{4}},
\ee
\be\label{h2}
H_{2}={x_{1}\over\gamma_{1}} +
{x_{2}\over\gamma_{2}}+
{x_{3}\over\gamma_{3}} +
{x_{4}\over\gamma_{4}}.
\ee
{} For the integrability of the system (\ref{system}), we  need one more
  integral of motion, $H_{3}$.
To find that integral let us take Nambu's form of the system
(\ref{system})
\be\label{ecu}
{\dot x_{n}}=\{x_{n},H_{1},H_{2},H_{3}\}=
\gamma_{1}\gamma_{2}\gamma_{3}\gamma_{4}\epsilon_{nmkl}
{\partial H_{1}\over\partial x_{m}}
{\partial H_{2}\over\partial x_{k}}
{\partial H_{3}\over\partial x_{l}}.
\ee
We found from (\ref{ecu}) a solution for $H_{3}$
\be\label{h3}
H_{3}=-{1\over 2}( {x_{1}\over\gamma_{1}}-
{x_{2}\over\gamma_{2}}+
{x_{3}\over\gamma_{3}}-
{x_{4}\over\gamma_{4}}).
\ee

Because we already have three integrals of motion, we can integrate the system
(\ref{system}). From (\ref{h2}) and (\ref{h3}) we get
\ba\label{x4}
x_{4}=\gamma_{4}({{H_{2}+2H_{3}}\over 2}-{x_{2}\over\gamma_{2}}), \cr
x_{3}=\gamma_{3}({{H_{2}-2H_{3}}\over 2}-{x_{1}\over\gamma_{1}})
\ea
and  (\ref{h1}) gives us
\be
{e^{x_{1}}\over\gamma_{1}}+{e^{x_{2}}\over\gamma_{2}}+
e^{-x_{1}{\gamma_{3}\over\gamma_{1}}}{e^{{\gamma_{3}\over
2}{(H_{2}-2H_{3})}}\over\gamma_{3}} +
e^{-x_{2}{\gamma_{4}\over\gamma_{2}}}{e^{{\gamma_{4}\over
2}{(H_{2}+2H_{3})}}\over\gamma_{4}}=H_{1}. \ee

So $x_{2}$ is an implicit function of $x_{1}$,
$x_{2}=n_{1}(x_{1},H_{1},H_{2},H_{3})$.
When
\be\label{gam}
{\gamma_{4}\over\gamma_{2}}=\pm1,\pm2,\pm3,-4,
\ee
 the function $n_{1}$ reduces to the composition of the elementary
functions.
When
\be\label{gam1}
{\gamma_{3}\over\gamma_{1}}=\pm1,\pm2,\pm3,-4,
\ee
 we have $x_1$ as a superposition of elementary functions
of $x_2$.
Similarly we can consider the cases for the ratios $\gamma_{3}\over\gamma_2$
and $\gamma_{4}\over\gamma_1$.

Now we can solve the equation for $x_{1}$,

\be
{\dot x_{1}}=\gamma_{1}(e^{x_{2}}-e^{x_{4}})=n_{2}(x_{1}),
\ee
by one quadrature,
\be
N(x_{1})=\int_{0}^{x_{1}}{dx\over n_{2}(x)}=t-t_{0}.
\ee
{\bf 5.} Conclusions\\

 As is well known, Nambu mechanics is a
 generalization of classical Hamiltonian mechanics introduced by
Yoichiro Nambu, \cite{nambu}.
In \cite{cohen,cate} it was demonstrated that several
Hamiltonian systems possessing dynamical symmetries can be realized in
the Nambu formalism of generalized mechanics.\\

In this paper we invented the system (\ref{din}) and
investigate the integrability properties of the
particular cases of the system by elementary methods
using Nambu--Poisson reformulation of Hamiltonian mechanics.

For the general case
we have two integrals of motion  for the system (\ref{din})

\be\label{h1n}
H_{1}=\sum_{n=1}^{N}{e^{x_{n}}\over\gamma_{n}},
\ee
\be\label{h2n}
H_{2}=\sum_{n=1}^{N}{x_{n}\over\gamma_{n}}.
\ee

For even N, $N=2M$, we found a third integral of motion

\be\label{h3n}
H_{3}={1\over 2}\sum_{n=1}^{2M}{(-1)^{n}x_{n}\over\gamma_{n}},
\ee
 but when $N\ge 5$, for integrability, we need extra integrals of motion.
The integrability properties of the system (\ref{din})
in the general case are under investigation, \cite{new}.

\end{document}